\documentclass[12pt,a4paper,final]{iopart}

\usepackage{iopams}  
\usepackage{graphicx}
\usepackage{units}
\usepackage[breaklinks=true,colorlinks=true,linkcolor=blue,urlcolor=blue,citecolor=blue]{hyperref}

\begin{document}

\title{Phase- and intensity-resolved measurements of above threshold ionization by few-cycle pulses}

\author{M. K\"ubel$^1$, M. Arbeiter$^3$,  C. Burger$^1,2$, Nora G. Kling$^1$, T. Pischke$^1$, R. Moshammer$^4$, T. Fennel$^{3,5}$, M.F. Kling$^{1,2}$, B. Bergues$^{1,2}$}
\address{$^1$ Department f\"ur Physik, Ludwig-Maximilians-Universit\"at, D-85748 Garching, Germany}
\address{$^2$ Max-Planck-Institut f\"ur Quantenoptik, D-85748 Garching, Germany}
\address{$^3$ Institute of Physics, University of Rostock, Universit\"{a}tsplatz 3, D-18051 Rostock, Germany}
\address{$^4$ Max-Planck-Institut f\"ur Kernphysik, D-69117 Heidelberg, Germany}
\address{$^5$ Max-Born-Institut, Max-Born-Stra\ss e 2A, D-12489 Berlin, Germany}
\eads{\mailto{matthias.kuebel@physik.uni-muenchen.de} \\ 
\mailto{thomas.fennel@uni-rostock.de}\\
\mailto{boris.bergues@mpq.mpg.de}
}

\begin{abstract}
We investigate the carrier-envelope phase and intensity dependence of the longitudinal momentum distribution of photoelectrons resulting from above-threshold ionization of argon by few-cycle laser pulses. The intensity of the pulses with a center wavelength of 750\,nm is varied in a range between $0.7 \times 10^{14}$ and $\unit[5.5 \times 10^{14}]{W/cm^2}$. Our measurements reveal a prominent maximum in the carrier-envelope phase-dependent asymmetry at photoelectron energies of 2\,$U_\mathrm{P}$ ($U_\mathrm{P}$ being the ponderomotive potential), that is persistent over the entire intensity range. Further local maxima are observed at 0.3 and 0.8\,$U_\mathrm{P}$. The experimental results are in good agreement with theoretical results obtained by solving the three-dimensional time-dependent Schr\"{o}dinger equation. We show that for few-cycle pulses, the amplitude of the carrier-envelope phase-dependent asymmetry provides a reliable measure for the peak intensity on target. Moreover, the measured asymmetry amplitude exhibits an intensity-dependent interference structure at low photoelectron energy, which could be used to benchmark model potentials for complex atoms.

\end{abstract}

\vspace{2pc}
\submitto{\JPB}
%\maketitle

\section{Introduction}

Above threshold ionization (ATI) of an atomic or molecular target is the absorption of multiple photons in excess of the ionization energy from a short and intense laser pulse \cite{Agostini1979,Becker2002,Milosevic2006}. ATI is at the heart of many phenomena involving the interaction of matter with strong laser fields and has led to several applications including high-harmonic generation \cite{Ferray1988} and attosecond metrology \cite{Hentschel2001,Krausz2009}. As such, ATI photoelectron spectra have been attracting a lot of attention over the past four decades, see e.g.~\cite{Agostini1979,Freeman1987,Corkum1989, Schafer1993, Paulus1994, Grasbon2003, Kling2008, Blaga2009, Bergues2011}. They have been utilized to measure physical quantities such as photoionization time delays \cite{Eckle2008, Pfeiffer2011}, or molecular orbitals \cite{Meckel2008} and dynamics \cite{Blaga2012, Boguslavskiy2012, Skruszewicz2015, Wolter2016}. Meanwhile, the different characteristic features that ATI spectra exhibit at high \cite{Paulus1994}, medium \cite{Paulus2001, Bergues2007, Gopal2009}, and low \cite{Blaga2009, Wu2012, Wolter2015} energies, have been investigated in numerous studies, and their origin could be traced back to either purely quantum or classical dynamics \cite{Becker2002,VPopruzhenko2014}.  

Due to the highly nonlinear nature of ATI, attosecond measurements in the strong-field regime tend to be sensitive to the actual waveform of the electric field and thus require an accurate knowledge of the laser pulse parameters on target, such as pulse duration, carrier-envelope phase (CEP), and peak intensity. So far, however, the accurate determination of the laser intensity on target by measuring the spatial beam properties often remains a challenging and laborious task. An alternative approach to tackle this problem is to use the high sensitivity of ATI to our advantage and utilize the information contained in the photoelectron spectra to determine the laser parameters. The approach of harvesting the information contained in the ATI spectra also allows the measurement of the CEP \cite{Paulus2003a, Rathje2012} and pulse duration \cite{Sayler2011a} of few-cycle pulses.  

%Several methods are based on evaluating the shape of 
The ATI spectra contain information on the laser intensity: ignoring the interaction of the released electron with the ionic core, a linearly polarized laser field with electric field amplitude $E_0$ and frequency $\omega$ can accelerate the electron to a kinetic energy of up to twice the ponderomotive potential (atomic units are used throughout)
\begin{equation}
    U_\mathrm{P} = E_0^2 / 4\omega^2. 
\label{Up}
\end{equation}
Recollision with the parent ion \cite{corkum1993,Schafer1993} and subsequent back-scattering enables the electron to reach energies up to 10\,$U_\mathrm{P}$ \cite{Paulus1994a}. Although the 2\,$U_\mathrm{P}$ or 10\,$U_\mathrm{P}$ cut-off energies enable measuring the peak electric field strength, their identification in the spectra is often somewhat ambiguous and yields a relatively large uncertainty of typically 20\%. Evaluating the ATI spectra for circular polarization \cite{Alnaser2004} removes some of the ambiguity, as all electrons reach an energy of $U_\mathrm{P}$\footnote{Note that in circularly polarized fields, $U_P$ has twice the value given in equation \ref{Up}.}. Still, generating perfectly circularly polarized fields is challenging in the few-cycle regime. Recently, a transferable intensity standard based on ionization of atomic hydrogen has been published \cite{Wallace2016}. However, the method, which consists in the measurement of the intensity dependent ionization yield from noble gases, requires a separate intensity scan. A similar scheme based on the referencing to hydrogen has also been demonstrated for measuring the absolute CEP \cite{Sayler2015}. 

Here, we systematically study the combined intensity and CEP dependence of the longitudinal momentum distribution of ATI of argon over almost one order of magnitude of intensity. We show that the CEP dependent momentum distribution allows for an unambiguous, precise and accurate intensity determination without the need to perform an intensity scan.

\section{Experimental Approach}

\begin{figure}
    \centering
    \includegraphics[width=0.95\textwidth]{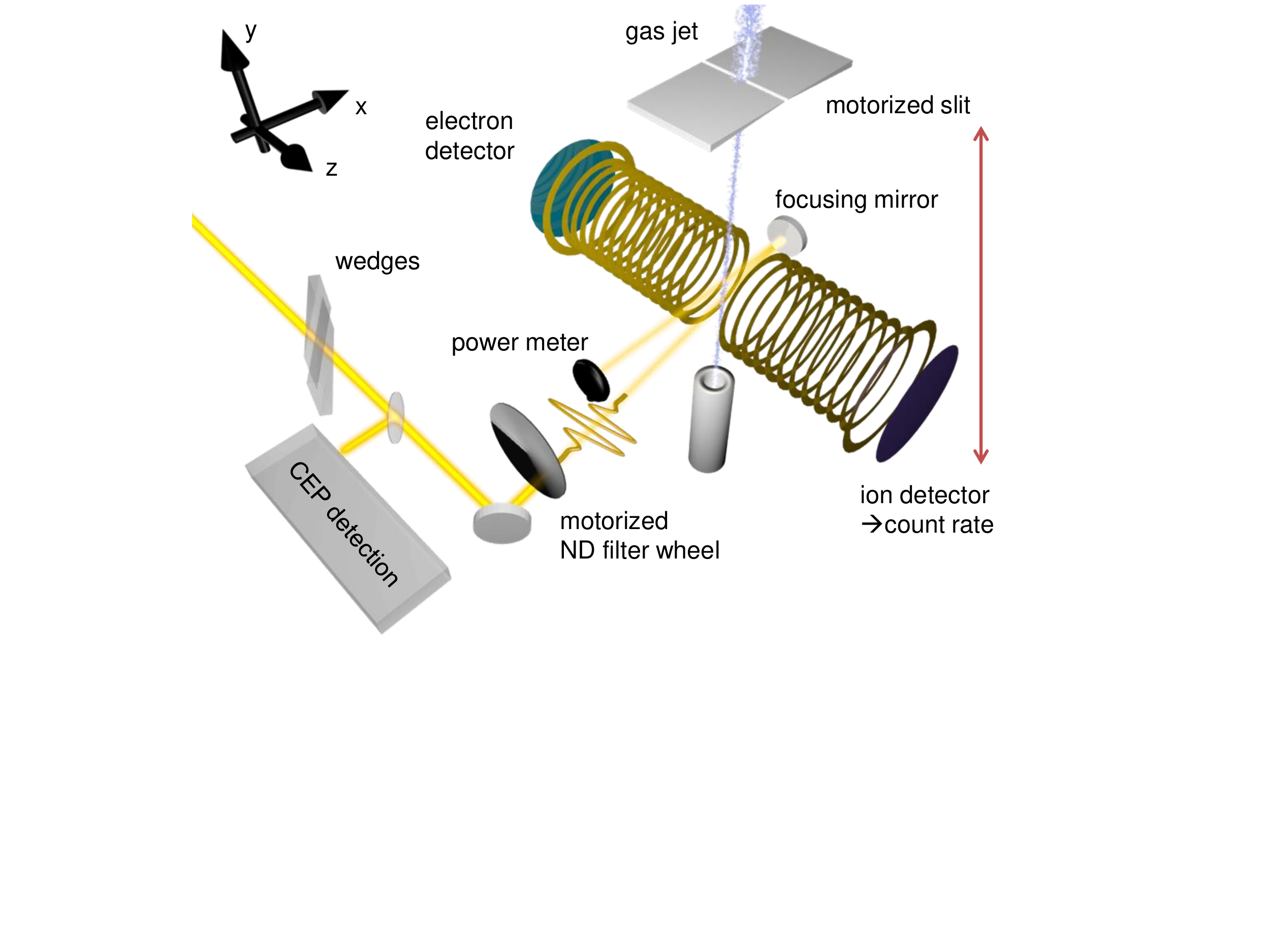}
    \vspace{-4cm}
    \caption{Schematic of the experimental setup. A small fraction of the few-cycle pulses is sent into an f-2f interferometer for CEP detection. The main part passes through a motorized neutral density (ND) filter wheel and is focused ($f=175$\,mm) into a Reaction Microscope (REMI) where photoelectron and photoion momentum distributions arising from strong-field ionization of argon are recorded. The thickness of the cold gas jet along $x$ can be adjusted using a motorized slit. The arrow on the right side represents a feedback from the ion count rate to the motorized slit, which allows for controlling the count rate over approximately two orders of magnitude. The laser power is recorded with an electronic power meter at the output of the REMI.}
    \label{Fig1}
\end{figure}
Carrier-envelope phase stable few-cycle laser pulses with a duration of 4.5\,fs (full width at half maximum of the intensity envelope) and a center wavelength of 750\,nm are obtained from a titanium:sapphire chirped pulse amplifier (Spectra Physics Femtopower HR CEP4) equipped with a gas-filled hollow-core fiber for spectral broadening. After pulse compression using chirped mirrors and fused silica wedges, the beam is sent into the beam path shown in Fig.~\ref{Fig1}.

An $f-2f$ interferometer is used to measure the CEP up to a constant offset value. For a quantitative study of the intensity dependence of strong-field ionization, it is important to vary the focal intensity in a well-controlled manner, i.e. without affecting the pulse duration or the focal intensity distribution. To achieve this, we use a neutral density filter wheel (Inconel NiCrFe coating) with an angle dependent transmission. The transmitted pulse energy and, hence, the intensity 
\begin{equation}
    I = \frac{P}{k}, \hspace{1cm} k = A f_\mathrm{rep} \tau = \mathrm{const},
    \label{eq:intensity}
\end{equation}
is varied by rotating the filter wheel without changing the focal cross section $A$ or pulse duration $\tau$. Here,  $f_\mathrm{rep} = 9900\,\mathrm{Hz}$ is the laser repetition rate, and $P$ the average laser power, which is recorded with an electronic power meter at the exit of the REMI. 
%The width of the motorized slit cutting into the gas jet is constantly adjusted to maintain a count rate of approximately 0.1 ions per laser pulse in the REMI while varying the laser power. 
In order to avoid false coincidences, the count rate is kept at approximately $0.1$ to $0.15$ ions per laser shot by reducing the target thickness when the intensity is increased. This is implemented via a feedback system that controls the width of the motorized slit cutting into the gas jet.

In order to cover a large intensity range, we performed two separate experiments using focal lengths of $f_1=17.5\,\mathrm{cm}$ and $f_2=10.0\,\mathrm{cm}$ for the low and high intensity regions, respectively.

\section{Results}

\begin{figure}
    \centering
    \includegraphics[width=\textwidth]{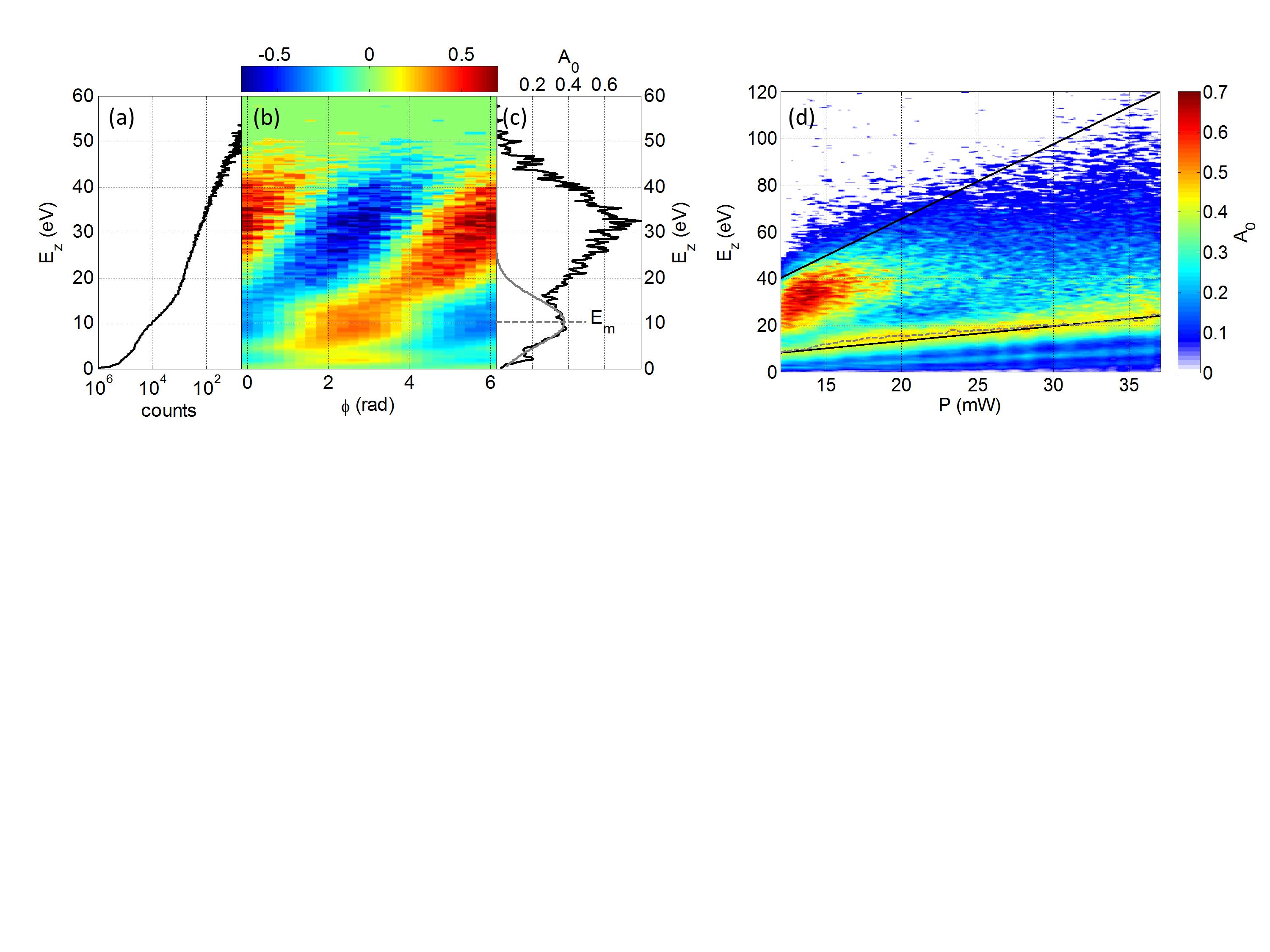}

   \vspace{-6.5cm}
    \caption{(a) Measured $E_z$-distribution of Ar$^+$ ions, where $E_z=p_z^2/2$, and $p_z$ is the momentum component along the laser polarization, for an average laser power of $P = 13.0\,\mathrm{mW}$ (b) Recorded asymmetry parameter as a function of $E_z$ and CEP. (c) Amplitude $A_0(E_z)$ of the CEP-dependent asymmetry parameter. The position $E_m$ of the local maximum $A_0(E_z)$ near 10\,eV is obtained from a Gaussian fit to $A_0(E_z)$. (d)  Measured asymmetry amplitude as a function of $E_z$ and average laser power $P$. The black solid lines represent estimates for $10\,U_\mathrm{P}$, and $2\,U_\mathrm{P}$, respectively.  The gray dashed line shows $E_m$ as a function of the average laser power.}
    \label{Fig2}
\end{figure}
The experimental results are summarized in Fig.~\ref{Fig2}. In panel (a), the number of detected Ar$^+$ ions recorded at a power $P = (13.0 \pm 0.3)\,\mathrm{mW}$ is plotted as a function of $E_z=p_z^2/2$, where $p_z$ is the ion momentum component along the laser polarization. Note that, due to momentum conservation, the ion momentum has the same magnitude as the photoelectron momentum. 

Figure \ref{Fig2}(b) shows the CEP-dependent asymmetry in the directional ion yield as a function of $E_z$ and CEP $\phi$. The CEP-dependent asymmetry $A$ is defined as

\begin{equation}
    A(E_z,\phi) = \frac{N(|p_z|,\phi) - N(-|p_z|,\phi)}{N(|p_z|,\phi)+N(-|p_z|,\phi)+\epsilon},
\end{equation}
where $N(|p_z|,\phi)$ and $N(-|p_z|,\phi)$ refer to the number of ions emitted with momentum $|p_z|$ and $-|p_z|$, respectively, and $\epsilon = 1$ is used to avoid division by zero. 

For each value of $E_z$, we extract the amplitude $A_0 (E_z)$ of the CEP-dependent asymmetry via Fourier transform of $A(E_z,\phi)$ and analysis of the $2\pi$ periodic component.
The obtained asymmetry amplitude $A_0 (E_z)$, which is plotted in Fig.~\ref{Fig2}(c), exhibits several local maxima. Besides the large maximum at energies in the recollision plateau (above 20\,eV), two other local maxima are apparent around 10\,eV and 2\,eV. The position of the 10-eV maximum, denoted as $E_m$ in the following, is obtained from a Gaussian fit to $A_0 (E_z)$ in the region around this peak.

In Fig.~\ref{Fig2}(d), the measured $E_z$-dependent asymmetry amplitude is displayed as a function of the laser power. Notably, the position of the low energy maxima scale linearly with the laser power. The upper black line indicates the position of the high energy ($10\,U_\mathrm{P}$) cutoff, estimated at low intensity and extrapolated to higher intensity values. The lower black line with a five times smaller slope indicates the estimated position of the $2\,U_\mathrm{P}$ cutoff. The gray line denotes the positions of the local asymmetry maximum, $E_m$, obtained from repeating the procedure outlined above for each recorded laser power. Interestingly, $E_m(P)$ coincides with the black line at $2\,U_\mathrm{P}$.

To investigate the relationship between $E_m$ and $2\,Up$, we numerically solved the three-dimensional time-dependent Schr\"{o}dinger equation (3D TDSE) using Qprop \cite{Bauer2006,Mosert2016}. The wavefunction of the initial $3p$ state has been obtained by imaginary time propagation. The effective potential employed in the theoretical description in single-active electron approximation \cite{Muller1999,Sayler2015} has the form
\begin{equation}
V_\mathrm{eff}=-\frac{1}{r}  - \frac{Z_\mathrm{eff}}{r} \exp \left(-\alpha r \right)
\end{equation}

and ensures the correct long-range $1/r$-behavior. The effective charge $Z_\mathrm{eff}=12.58$ and the screening length $\alpha=\unit[1.5]{a.u.}$ have been obtained by matching the ionization potential ($I_\mathrm{P}$) to the correct value for argon and by  minimizing the deviation of the $3p$ wavefunction from the corresponding prediction of an atomic all-electron Dirac-LDA code. The spectra of the time-dependent calculations are extracted using a window operator method and then analyzed analogously to the experimental data. For the simulation, a laser wavelength of 750\,nm, and a pulse duration of $4.5\,\mathrm{fs}$ (full width at half-maximum of the intensity envelope) are used. The results are averaged over the intensity distribution resulting from a Gaussian beam profile neglecting the variation of the beam waist along the laser propagation, if not otherwise specified.

\begin{figure}
\centering
\includegraphics[width=0.9\textwidth]{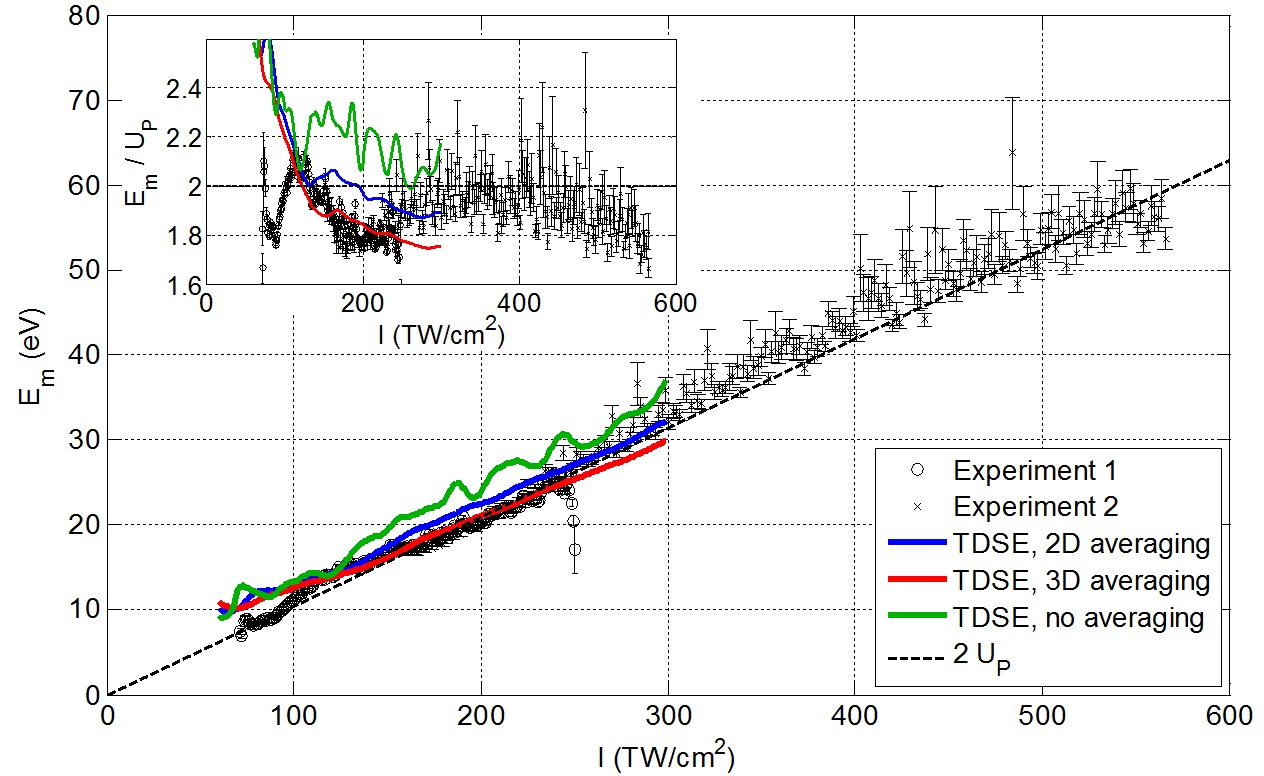}
%\vspace{-2.2cm}
\caption{Measured and calculated intensity dependence of the position $E_m$ of the asymmetry maximum. Shown are the results of two separate experiments, covering different intensity ranges. Theoretical results are shown for focal volume averaging with constant beam waist (2D), and varying beam waist (3D), and without focal volume averaging. The dashed black line indicates $2\,U_\mathrm{P}$. The inset shows $E_m$ in units of $U_\mathrm{P}$.}
\label{Fig3}
\end{figure}

The measured and calculated intensity dependence of the position of the local asymmetry maximum $E_m$ are compared in Fig.~\ref{Fig3}. For the calibration of the experimental intensity, only the proportionality constant between focal intensity and recorded laser power needs to be adjusted (see equation \ref{eq:intensity}). 
We find $k_1 = 0.1558\,\mathrm{mW}/(\mathrm{TW/cm}^{2})$ for experiment 1 and 
$k_2 = 0.0424\,\mathrm{mW}/(\mathrm{TW/cm}^{2})$ for experiment 2. 
Over the entire range of the two measurements, $E_m(I)$ remains close to the $2\,U_\mathrm{P}$ line. 
As shown in the inset, the deviation is always smaller than 20\% and for most intensities smaller than 10\%.

\section{Discussion}
For intensities above $9 \times 10^{13}\,\mathrm{W/cm}^2$, the theoretical results confirm the empirical observation that $E_m$ is associated with the direct-electron cut-off at $2\,U_\mathrm{P}$. Thus, the measurement of $E_m$ yields an estimate for the peak intensity given by: 
\begin{equation}
    I \approx (E_m / 12\,\mathrm{eV}) (\lambda / 800\,\mathrm{nm}) 10^{14}\,\mathrm{W/cm}^2.
\end{equation}
The precision of the experimental intensity determination decreases for high intensities where the asymmetry maximum at $E_m$ becomes broader. The data presented in Fig.~\ref{Fig3} contains approximately 1 million counts for each intensity value, more than 400 million counts in total, acquired over a period of 120 hours. 

To test the robustness of the intensity calibration with respect to the focal intensity distribution, we average the calculated spectra over the focal volume for two different scenarios. In the first scenario, the signal is averaged over the focal plane (i.e. over two dimensions), thus neglecting the variation of the beam waist along the laser propagation direction. This represents a good approximation when the extension of the gas target along the laser propagation direction is much smaller than the Rayleigh range. In the second scenario, we take the variation of the beam waist along the laser propagation into account, and integrate over the full 3D intensity distribution in the focus of a Gaussian beam. This represents a good approximation when the extension of the gas target along the laser propagation direction is larger than the Rayleigh range. The results of these two calculations are compared to those obtained at the peak intensity only. As can be seen in Fig. 3, the intensity averaging has two effects. First, it washes out modulations visible in the fixed-intensity results (green line). Second, it reduces the effective intensity of the laser focus, as intensities lower than the peak value contribute to the intensity-averaged yield. In the 3D case, the contribution of lower intensities is larger than in the 2D case, resulting in a lower effective intensity. The difference between 2D and 3D averaging amounts to approximately 10\% and can be considered a good measure for typical uncertainties of the focal shape. The transition from 2D to 3D focal volume averaging as the size of the gas jet is increased with respect to the size of the laser focus has been discussed in detail in Ref.~\cite{Kubel2016a}. 

At intensities below 100\,TW/cm$^2$, the TDSE results deviate from the experimental results. We attribute these deviations to inaccuracies of the short-range behavior of the effective potential used for the TDSE. The deviations also suggest that slow electrons may be particularly sensitive to the effective potential landscape, which is, in turn, particularly hard to model close to the ionic core, where electron correlations play a more important role.

\begin{figure}
    \centering
    \includegraphics[width=\textwidth]{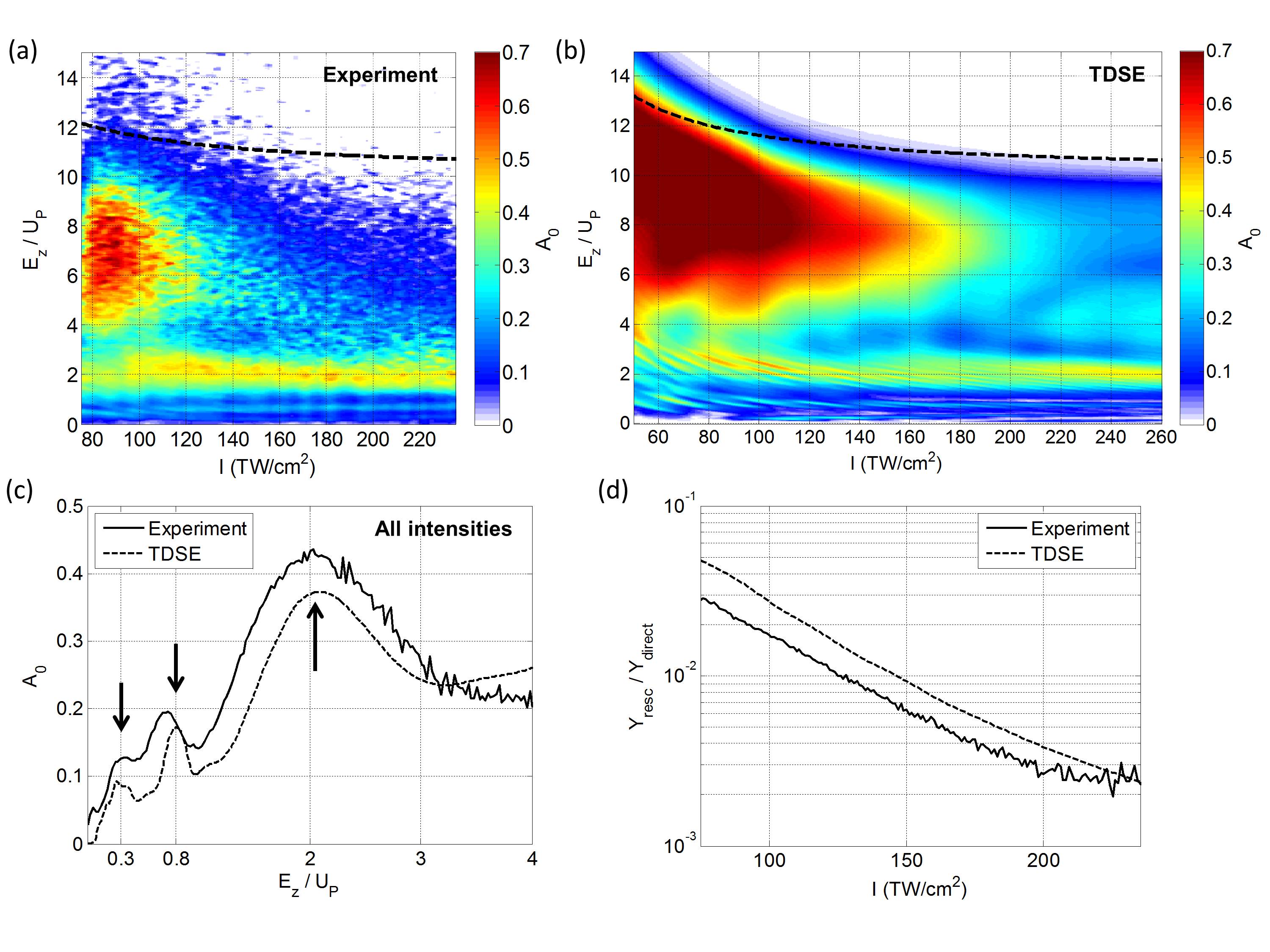}
    \vspace{-1.2cm}
    \caption{Measured (a) and calculated (b) asymmetry amplitude as a function of $E_z$ and intensity, where $E_z$ is expressed in units of $U_\mathrm{P}$. The black dashed line represents the cut-off law given in Ref.~\cite{Busuladzic2006}. (c) Asymmetry amplitude as a function of $E_z$ in units of $U_\mathrm{P}$, integrated over all intensities. The position of the local maxima of the asymmetry amplitude below $2\,U_\mathrm{P}$ are marked with arrows. (d) Calculated and measured ratio of the yields for rescattered (with $4\,U_\mathrm{P} < E_z < 10\,U_\mathrm{P}$) and direct electrons (with $0.5\,U_\mathrm{P} < E_z < 2\,U_\mathrm{P}$).}
    \label{Fig4}
\end{figure}

The accurate determination of the experimental intensity on target facilitates the detailed discussion of the combined intensity and CEP dependence of the photoelectron spectra. To this end, we plot in Fig.~\ref{Fig4} the measured and predicted asymmetry amplitudes as a function of intensity and $E_z$, in units of $U_\mathrm{P}$.

We first discuss the behavior of the region of rescattered electrons, $E > 4\,U_\mathrm{P}$. As can be seen in Figs.~\ref{Fig4} (a) and (b), the high energy cut-off decreases with increasing intensity. In the TDSE results, this decrease is rather consistent with the formula for the cut-off energy $E_c = 10\,U_\mathrm{P} + 0.538\,I_\mathrm{p}$ given in Ref.~\cite{Busuladzic2006}, which is represented by the dashed black line. We attribute the faster decrease of the asymmetry amplitude in the experimental data to the limited signal-to-noise ratio. As shown in Fig.~\ref{Fig4}(d), the ratio of recollision and direct electrons signal indeed decreases with increasing intensity, which can be understood as a consequence of the decreasing rescattering probability with increasing electron energy.      
Another prominent feature in Fig.~\ref{Fig4}(a) and \ref{Fig4}(b) is the decrease of the asymmetry amplitude in the rescattering region at higher intensity. This may be due to the onset of saturation of the ionization probability, which washes out the contrast of ionization probability for half-cycles with slightly different field strengths.

\begin{figure}
    \centering
    \includegraphics[width=\textwidth]{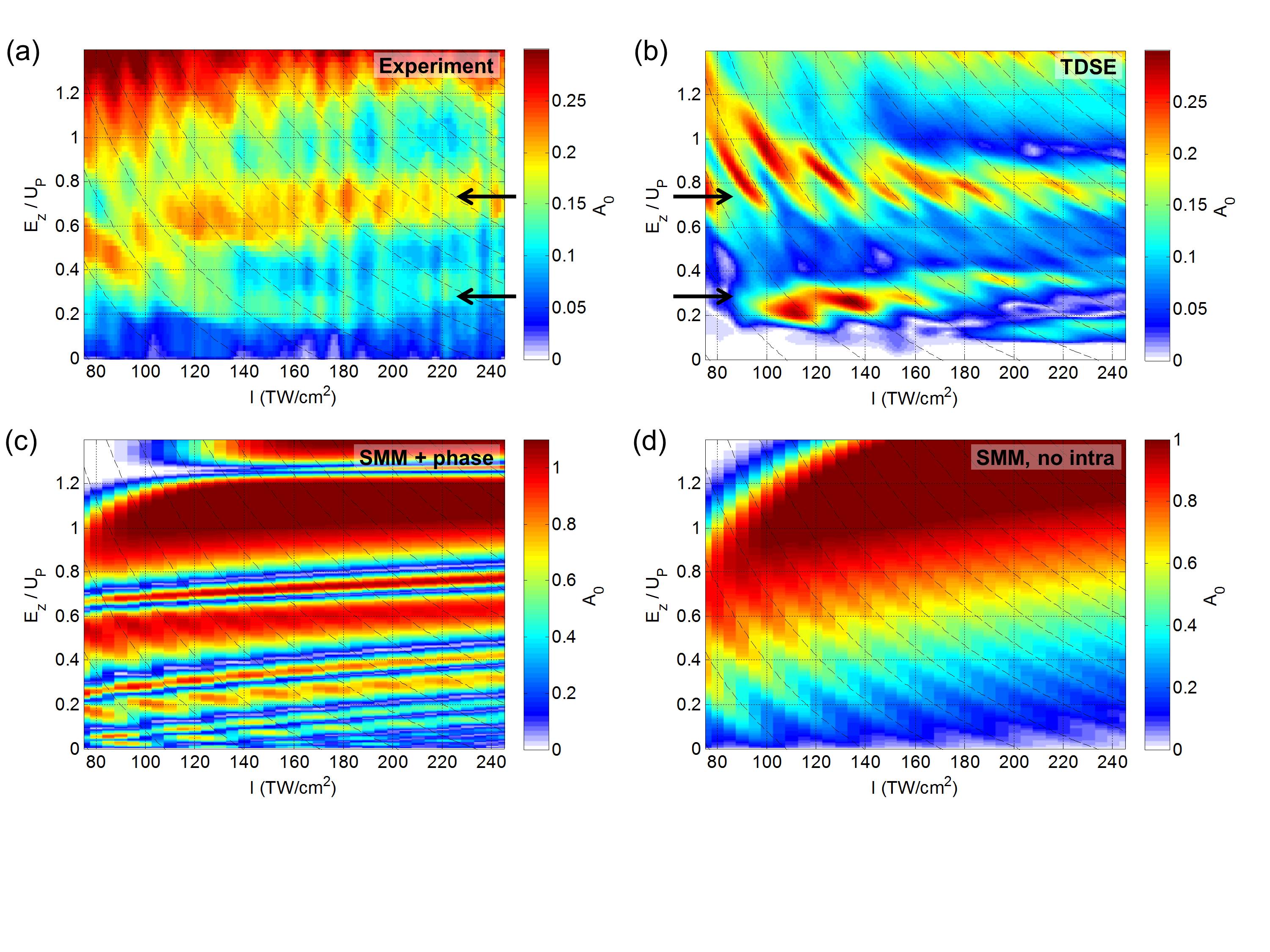}
    \vspace{-2.5cm}
    \caption{Close up view of the measured (a) and simulated (b-d) asymmetry amplitude at low energies. The arrows mark the position of the features highlighted in Fig.~\ref{Fig4} (c). The dashed lines mark the expected peak positions of an ATI comb with a spacing corresponding to the laser wavelength of 750\,nm, taking the ponderomotive shift of the peaks into account. Panel (b) shows results of the 3D TDSE. Panel (c) shows results for the extended simple man's model (SMM) with interferences. In panel (d), intracycle interferences are artificially  switched off.}
    \label{Fig5}
\end{figure}

We now turn towards the region of direct electrons $E \le 2\,U_\mathrm{P}$. The most prominent feature is the large horizontal bar at $2\,U_\mathrm{P}$. We further observe two additional local asymmetry maxima, as indicated by the arrows in Fig.~\ref{Fig4}(c). The maximum at $0.8\,U_\mathrm{P}$ corresponds to the maximum observed in Fig.~\ref{Fig2}(c) at $2\,\mathrm{eV}$. The maximum at $0.3\,U_\mathrm{P}$ emerges only at intensities above 120\,TW/cm$^2$ in the experimental data (see Fig.~\ref{Fig5}(a)). The observed maxima are attributed to intracycle interferences \cite{Gribakin1997, Kopold1999, Kopold2000a}. The latter result from the fact that there are two quantum trajectories of direct electrons ionized at two different instants within the same laser cycle, which lead to the same final momentum state. The intracycle interference structures, which were initially observed in above threshold detachment spectra of negative ions \cite{Reichle2001, Reichle2003}, were shown to govern the shape of the photoelectron spectra \cite{Bergues2007}. In the context of strong field ionization of atoms with few-cycle pulses, the intracycle interference effect leads to CEP dependent modulations of the ATI spectra \cite{Gopal2009}.

In order to support the claim that the observed maxima result from intracycle interferences, we have performed a 1D semi-classical simulation of interferences in the ATI spectra produced by 4.5\,fs, 750\,nm laser pulses of different intensities. For a given final momentum $p$, electron trajectories are launched at all points in time $t_n$ satisfying $A(t_n) = p$.
The phase of each trajectory is calculated by computing the action $S$ of a trajectory as follows

\begin{eqnarray}
S(t_n) &=& S_g(t_n) + S_a(t_n) + S_c(t_n), \\
S_g(t_n) &=& \int_{0}^{t_n} \, -I_\mathrm{P} \, dt, \\
S_a(t_n) &=& \mathrm{sgn}[-E(t_n)] \, \pi/2, \\
S_c(t_n) &=& \int_{t_n}^\infty \, [A(t)-A(t_n)]^2/2 \,dt. 
\end{eqnarray}
The atomic action $S_a$ takes into account the parity of the groundstate $p$ orbital.

The ionization probability for each trajectory $R(t_n)$ is calculated using the ionization rates given in Ref.~\cite{Tong2005}. The photoelectron spectrum is calculated as

\begin{equation}
W(p) = |\sum_n \sqrt{R(t_n)}\,\exp{\left(i S(t_n)\right)}|^2,
\label{smm_spec}
\end{equation}
The CEP-dependent asymmetry is obtained from the spectra following the same procedure as for the TDSE results. We present in Fig.~\ref{Fig5} the measured and calculated asymmetry amplitude maps in the range below $2\,U_\mathrm{P}$.

The asymmetry maxima highlighted in Fig.~\ref{Fig4} (c) are clearly visible and marked by arrows in the experimental and TDSE results shown in Figs.~\ref{Fig5}(a) and (b), respectively. A qualitatively similar series of nearly horizontal maxima can be seen in the results of the semi-classical calculations shown in Fig.~\ref{Fig5}(c), while it is absent in Fig.~\ref{Fig5}(d). There, intracycle interferences are artificially switched off by considering only one of the two interfering trajectories within each cycle, but allowing \emph{inter}cycle interferences. This indicates that the intracycle interference effect is indeed responsible for the observed horizontal series of asymmetry maxima. 

In addition to the intracycle interferences, the TDSE results exhibit strong periodic and intensity-dependent modulations that are most pronounced at low intensities. The comparison to the curved dashed lines in Fig.~\ref{Fig5} shows that the periodic maxima in the asymmetry amplitude can be associated with a series $(n = 10,~11,~12,~...)$ of ponderomotively shifted ATI peaks, arising from \emph{inter}cycle interferences, given by
\begin{equation}
 E_n / U_\mathrm{P} = (n \hbar \omega -I_\mathrm{P}) / U_\mathrm{P} - 1.
\end{equation}
The spacing of the ATI maxima are consistent with modulations visible in the experimental data (Fig.~\ref{Fig5}(a)) and semi-classical results (Fig.~\ref{Fig5}(c) and (d)).

We observe that the amplitude of the asymmetry is modulated by both, ATI peaks \cite{Abel2009} and intracycle interferences \cite{Arbo2010}.  Further investigations are needed to determine the reasons for the residual mismatch between TDSE results and experiment. As pointed out above, inaccuracies in the model potential used for argon in the 3D TDSE simulations most likely affect low-energy electrons. The observed features reflect the structure of the ionic potential. Thus, in principle, they could be used to benchmark model potentials for atomic or molecular targets. 

\section{Conclusion}

In conclusion, our systematic study of ATI of argon as a function of CEP and intensity in the few-cycle regime has revealed that the direct electron cut-off at $2\,U_\mathrm{P}$ is characterized by a pronounced maximum of the CEP-dependent asymmetry. This feature provides a convenient benchmark for a robust and accurate peak intensity determination of linearly polarized few-cycle pulses over a large intensity range, and without the need for performing a separate intensity scan. The rich structure observed in the CEP-dependent asymmetry amplitude of ATI  results from an interplay of ponderomotively shifted ATI peaks and intracycle interference.

\section*{Acknowledgements}
We acknowledge fruitful discussions with D. Bauer and T. Pfeifer. M.K., M.F.K., and B.B. acknowledge support from the Max Planck Society and the DFG via SPP1840, LMUexcellent, and the cluster of excellence "Munich Centre for Advanced Photonics" (MAP). We are grateful for support by the European Union via the ERC project ATTOCO and the H2020 project ATTOCHEM (657544). T.F. acknowledges support from the DFG via SFB652, SPP1840, and a Heisenberg fellowship. Computer time has been provided by the North German Supercomputing Alliance (HLRN) within project mvp00010.

\section*{References}

\bibliographystyle{unsrt}

%\section*{References}
\bibliography{ati_i_phi}
%\begin{thebibliography}{10}
%\bibitem{ref1} J.~Doe, Article name, \textit{Phys. Rev. Lett.}

%\bibitem{ref2} J.~Doe, J. Smith, Other article name, \textit{Phys. Rev. Lett.}

%\bibitem{web} \href{http://www.google.pl}{www.google.pl}
%\end{thebibliography}

\end{document}